# Dzyaloshinskii-Moriya interaction and magnetic skyrmions induced by curvature

Yonglong Ga,[1,2] Qirui Cui,[1] Jinghua Liang,[1] Dongxing Yu,[1] Yingmei Zhu,[1] Liming Wang[1] and Hongxin Yang[1,2,*]

[1]*Ningbo Institute of Materials Technology and Engineering, Chinese Academy of Sciences, Ningbo 315201, China*

[2]*Center of Materials Science and Optoelectronics Engineering, University of Chinese Academy of Sciences, Beijing 100049, China*

**Abstract**

Realizing sizeable Dzyaloshinskii-Moriya interaction (DMI) in intrinsic two-dimensional (2D) magnets without any manipulation will greatly enrich potential application of spintronics devices. The simplest and most desirable situation should be 2D magnets with intrinsic DMI and intrinsic chiral spin textures. Here, we propose to realize DMI by designing periodic ripple structures with different curvatures in low-dimensional magnets and demonstrate the concept in both one-dimensional (1D) $CrBr_2$ and two-dimensional (2D) $MnSe_2$ magnets by using first-principles calculations. We find that DMIs in curved $CrBr_2$ and $MnSe_2$ can be efficiently controlled by varying the size of curvature $c$, where $c$ is defined as the ratio between the height $h$ and the length $l$ of curved structure. Moreover, we unveil that the dependence of first-principles calculated DMI on size of curvature $c$ can be well described by the three-site Fert-Lévy model. At last, we uncover that field-free magnetic skyrmions can be realized in curved $MnSe_2$ by using atomistic spin model simulations based on first-principles calculated magnetic parameters. The work will open a new avenue for inducing DMI and chiral spin textures in simple 2D magnets via curvature.

*Introduction.*—The antisymmetric exchange coupling, Dzyaloshinskii-Moriya interaction (DMI) [1,2], arising from inversion symmetry breaking and spin-orbit coupling (SOC), plays an essential role in stabilizing topological protected non-collinear chiral magnetic configurations [3-7]. The presence of DMI is confirmed in the noncentrosymmetric B20 material MnSi [8], in which magnetic skyrmions are first observed. Recently, many reports have further demonstrated DMI can be induced at interfaces of $3d/5d$ metal [9-11], metal/oxides [12-15], or light elements such as graphene [16,17], oxygen and hydrogen, etc. [18,19]. Notably, experimental progresses have demonstrated that magnetic order of two-dimensional (2D) magnets can persist down to monolayer, which is useful for the study of fundamental physics and for engineering spintronic devices [20-22]. As for non-collinear magnetism, several effective methods are proposed to obtain sizeable DMI in 2D materials, such as in 2D Janus magnets [23-25], intrinsic 2D multiferroics [26,27] and van der Waals Ferromagnet-Based heterostructures, etc. [28-32], in which inversion symmetry breaking plays a significant role.

In practice, it is inevitable to induce ripples for a 2D material either free standing or on a substrate as long as the size of 2D material is large enough, which has been reported in graphene [33,34], and $MoS_2$ [35,36], etc. If such a curved system is magnetic, it is highly possible to achieve DMI in pure 2D magnets. Here, we propose to realize sizeable DMI by designing periodic ripple structures with different curvatures in low-dimensional magnets. We demonstrate the concept in both 1D $CrBr_2$ and 2D $MnSe_2$ structures, in which DMI can be efficiently tuned by magnitude of curvatures. Moreover, we validate that the variation of curvature-dependent DMI can be well interpreted by Fert-Lévy mechanism. Interestingly, using atomistic spin model simulations based on first-principles calculated magnetic parameters, we find that field-free chiral magnetic skyrmions can be realized in 2D $MnSe_2$. These examples provide a new routine towards realizing chiral magnetic skyrmions in 2D magnets with long-range magnetic orderings, e.g., $CrGeTe_2$ [37], $CrI_3$ [38] and $Fe_3GeTe_2$ [39], etc.

*Computational details.*—Our first-principles calculations are performed within the framework of density-functional theory (DFT) as implemented in the Vienna *ab initio* simulation package (VASP) [40]. We choose the Predev-Burke-Ernzerhof (PBE) functionals of generalized gradient approximation (GGA) [41] to deal with the exchange-correlation energy. Besides, Projector augmented plane wave (PAW) method [42,43] is adopted to deal with the interaction between nuclear electrons and valence electrons. In order to correctly describe the 3$d$ electrons, we employ the GGA+$U$ method [44] with an effective $U = 2$ eV and $U = 3$ eV for Mn and Cr as reported in previous studies [45,46], respectively. We build up a $7 \times 1 \times 1$ $CrBr_2$ and $6\sqrt{3} \times 1 \times 1$ $MnSe_2$ supercells to construct magnetic structures with different curvatures. The energy cutoff for plane wave expansion is set to 520 eV. Γ-centered $7 \times 1 \times 1$ for 1D $CrBr_2$ and $2 \times 18 \times 1$ for 2D $MnSe_2$ k-point meshes are adopted for the Brillouin zone (BZ) integration to relax supercell, respectively. All the structures are fully relaxed until the Hellmann-Feynman force acting on each atom and convergence criteria for energy are 0.001 eV/Å and is $1 \times 10^{-7}$ eV, respectively.

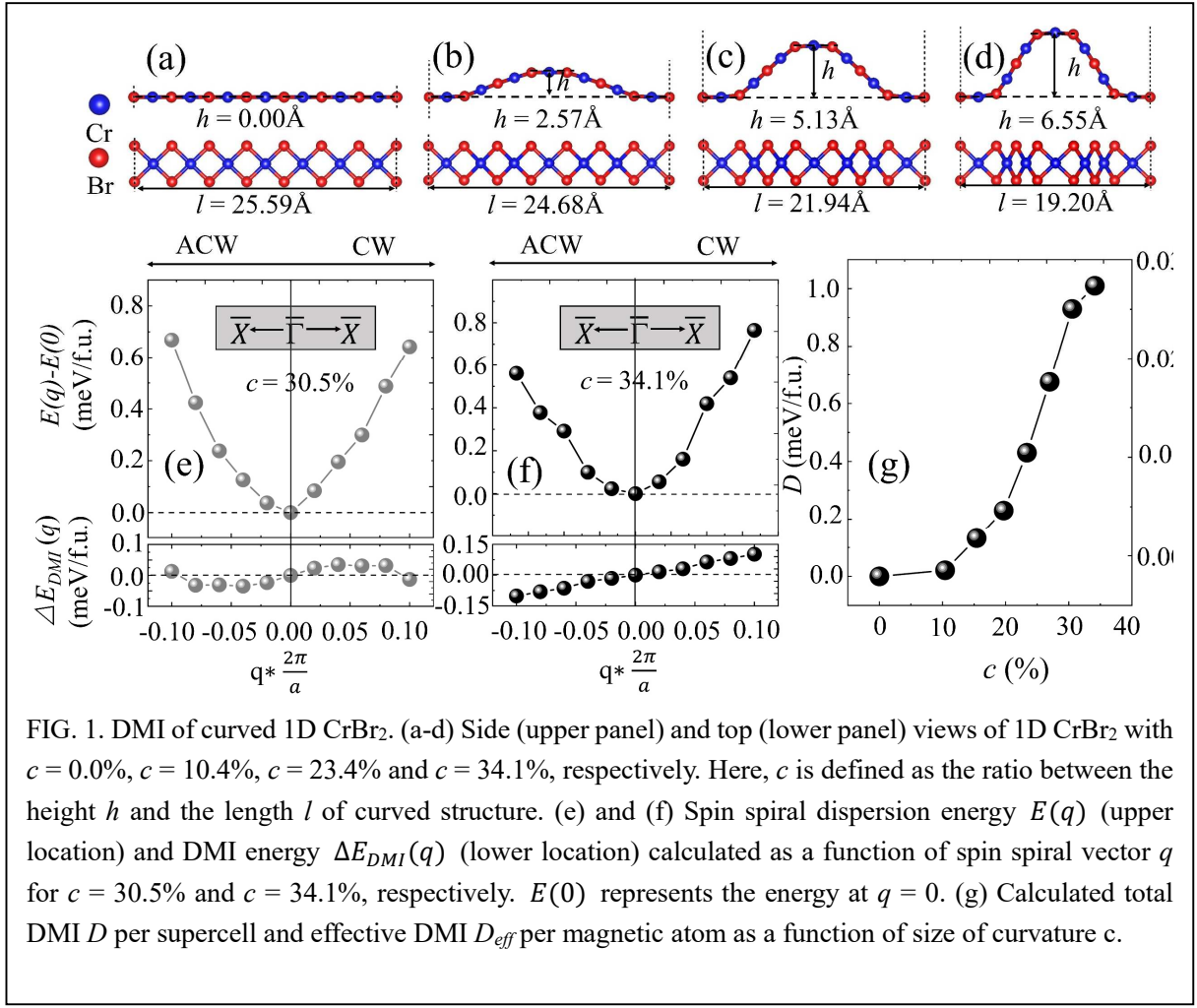

FIG. 1. DMI of curved 1D $CrBr_2$. (a-d) Side (upper panel) and top (lower panel) views of 1D $CrBr_2$ with $c = 0.0\%$, $c = 10.4\%$, $c = 23.4\%$ and $c = 34.1\%$, respectively. Here, $c$ is defined as the ratio between the height $h$ and the length $l$ of curved structure. (e) and (f) Spin spiral dispersion energy $E(q)$ (upper location) and DMI energy $\Delta E_{DMI}(q)$ (lower location) calculated as a function of spin spiral vector $q$ for $c = 30.5\%$ and $c = 34.1\%$, respectively. $E(0)$ represents the energy at $q = 0$. (g) Calculated total DMI $D$ per supercell and effective DMI $D_{eff}$ per magnetic atom as a function of size of curvature c.

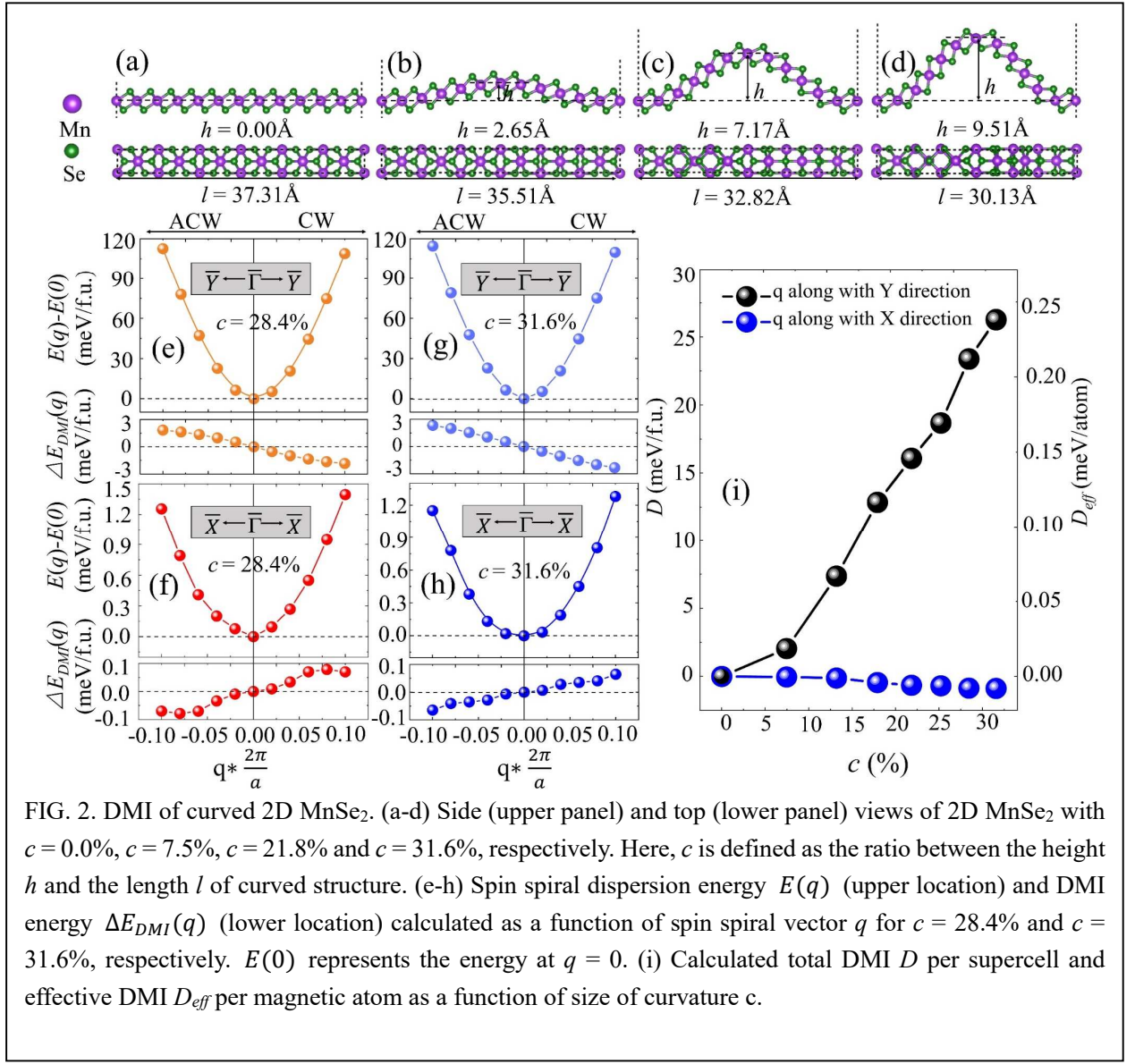

FIG. 2. DMI of curved 2D MnSe$_2$. (a-d) Side (upper panel) and top (lower panel) views of 2D MnSe$_2$ with $c = 0.0\%$, $c = 7.5\%$, $c = 21.8\%$ and $c = 31.6\%$, respectively. Here, $c$ is defined as the ratio between the height $h$ and the length $l$ of curved structure. (e-h) Spin spiral dispersion energy $E(q)$ (upper location) and DMI energy $\Delta E_{DMI}(q)$ (lower location) calculated as a function of spin spiral vector $q$ for $c = 28.4\%$ and $c = 31.6\%$, respectively. $E(0)$ represents the energy at $q = 0$. (i) Calculated total DMI $D$ per supercell and effective DMI $D_{eff}$ per magnetic atom as a function of size of curvature c.

*1-Dimensinal CrBr$_2$.*—For simplification of analysis, we first investigate curvature-dependent DMI in 1D magnet CrBr$_2$. Similar to CuCl$_2$ and CuBr$_2$, bulk CrBr$_2$ has the monoclinic space group C12/m1 [47-49]. In each CrBr$_2$ layer, the ribbons made up of edge-sharing CrBr$_4$ squares along b-axis. It has been not experimentally reported whether one-dimensional CrBr$_2$ can be exfoliated from bulk so far. In our paper, we use the simple 1D structure to demonstrate the dependence of DMI on size of curvature *c*. The obtained 1D CrBr$_2$ has D$_{2h}$ crystal symmetry which preserved spatial inversion symmetry, as shown in Fig. 1(a). For achieving ripples, a 7 × 1 × 1 supercell is constructed and the length *l* of entire supercell is adjusted gradually in our calculations. Next, we keep two unit cells (located at the edges of supercell) fixed in z direction and relax all other atoms to obtain final curved structures (Figs. 1(b)-1(d)). In order to articulate our ideas, we present the schematic diagram (Fig. 5) consisting of sine and piecewise functions, in which the inversion symmetry with symmetry center $i$ of sine function (red dotted line) will be broken in the piecewise function (blue line). This piecewise structure is an approximation for the effect of substrate on curvature. With this premise, we study how magnetic parameters change under different size of curvature $c$ for a series of curved structures (Figs. 1(a)-1(d)). Firstly, we study curvature-dependent magnetic properties of 1D $CrBr_2$. Figure 6(a) shows the ferromagnetic (FM) and antiferromagnetic (AFM) configurations. It is found that the calculated energy difference between AFM and FM evidently decreases with increasing $c$, as shown in Fig. 6(c). Moreover, the spin ordering tends to be AFM when $c$ is larger than 15.4%. In Fig. 7(a), we also investigate absolute value of magnetic moment per Cr atoms as $c$ increases. In general, the magnetic moment variation of each magnetic atom is relatively weak and tends to decrease with increase of curvature.

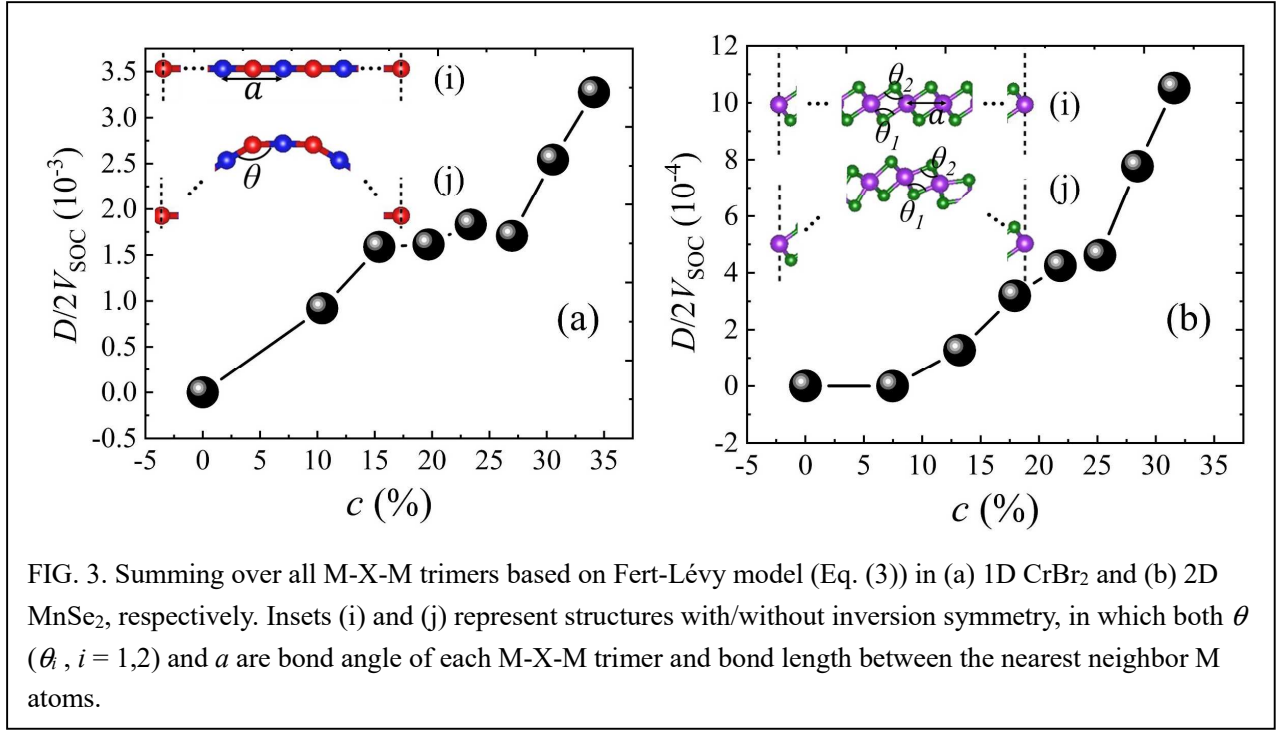

FIG. 3. Summing over all M-X-M trimers based on Fert-Lévy model (Eq. (3)) in (a) 1D $CrBr_2$ and (b) 2D $MnSe_2$, respectively. Insets (i) and (j) represent structures with/without inversion symmetry, in which both $\theta$ ($\theta_i$, $i = 1,2$) and $a$ are bond angle of each M-X-M trimer and bond length between the nearest neighbor M atoms.

Figure 1(g) shows the result of DMI for curved 1D $CrBr_2$ with different $c$. Obviously, DMI is strongly dependent on size of curvature $c$ and the calculated DMI strength increases with the increment of size of curvature $c$. Here, positive $D$ represents spin configuration of anticlockwise (ACW) chirality, negative one represents clockwise (CW) chirality. For total

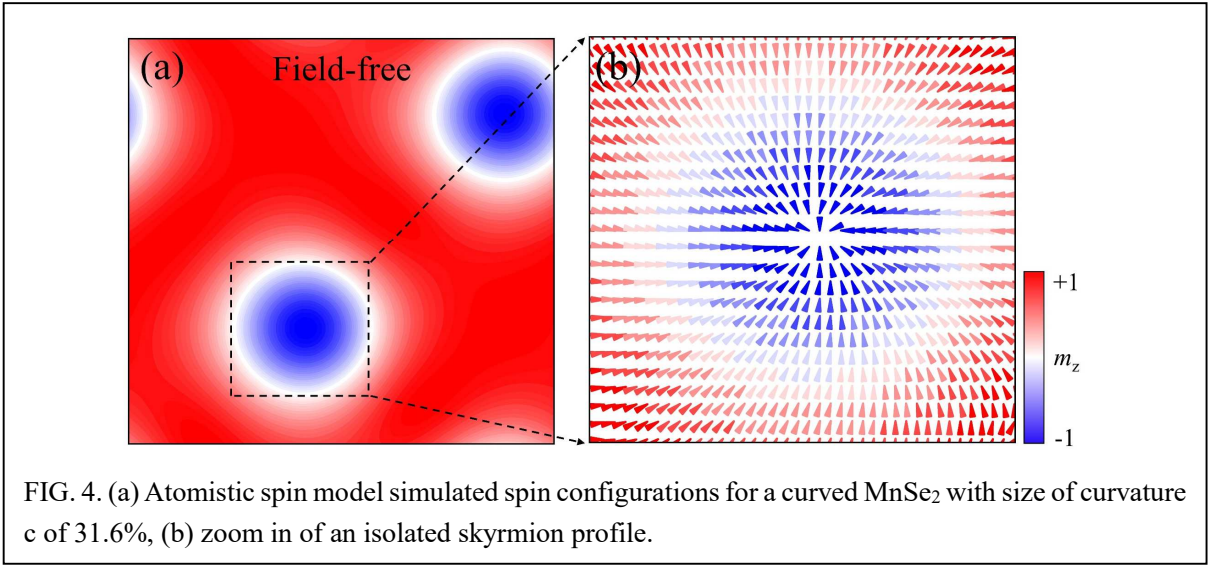

FIG. 4. (a) Atomistic spin model simulated spin configurations for a curved MnSe$_2$ with size of curvature c of 31.6%, (b) zoom in of an isolated skyrmion profile.

DMI $D$ of 1D CrBr$_2$ with different size of curvature $c$, its value varies from 0 to 1.0 meV/f.u.. Otherwise, In Figs. 1(e)-1(f) the total energy $E(q)$ and DMI energy $\Delta E_{DMI}(q)$ including SOC are presented as a function of spin spiral vector $q$ along $\overline{\Gamma} - \overline{X}$ direction in 2D BZ for $c$ = 30.5% and $c$ = 34.1%, respectively. The extracted DMI are found to be 0.93 and 1.00 meV/f.u., respectively, suggesting that DMI of 1D CrBr$_2$ is directly associated with the existence of curvature. Considering the fact that 2D magnets, such as CrGeTe$_3$, CrI$_3$, Fe$_3$GeTe$_2$, and MnSe$_2$ [50] with long-range ferromagnetic orderings have been synthesized experimentally, we take the simplest one MnSe$_2$ as an example to further investigate the feasibility of tunable DMI in 2D magnets.

*2-Dimensional MnSe$_2$.*—It is recently reported that monolayer 2D MnSe$_2$ with D$_{3d}$ crystalsymmetry is a room temperature ferromagnet [50]. The optimized MnSe$_2$ structure is presented in Fig. 2(a). It can be seen that each Mn atom is surrounded by six non-magnetic Se atoms that make up an octahedron. Similar to 1D CrBr$_2$, we construct a $6\sqrt{3} \times 1 \times 1$ supercell to obtain ripple structures, in which Mn atoms of three unit cells (located at initial and terminal of the supercell) are fixed in the z direction and the other atoms are relaxed (Figs. 2(b)-2(d)). To understand magnetism associated with curved structures, we construct FM and AFM states in Fig. 6(b). It is found that the energy difference between AFM and FM suddenly decreases and then keeps increasing as the $c$ gradually increases (Fig. 6d). We also compute the magnetic moment of Mn atoms from Mn1 to Mn12, as shown in Fig. 7d. Compared to 1D CrBr$_2$, there is a distinct difference that the magnetic moments of Mn atoms located at peaks and troughs have large variation due to the locally severe structural deformation, while others don't change much.

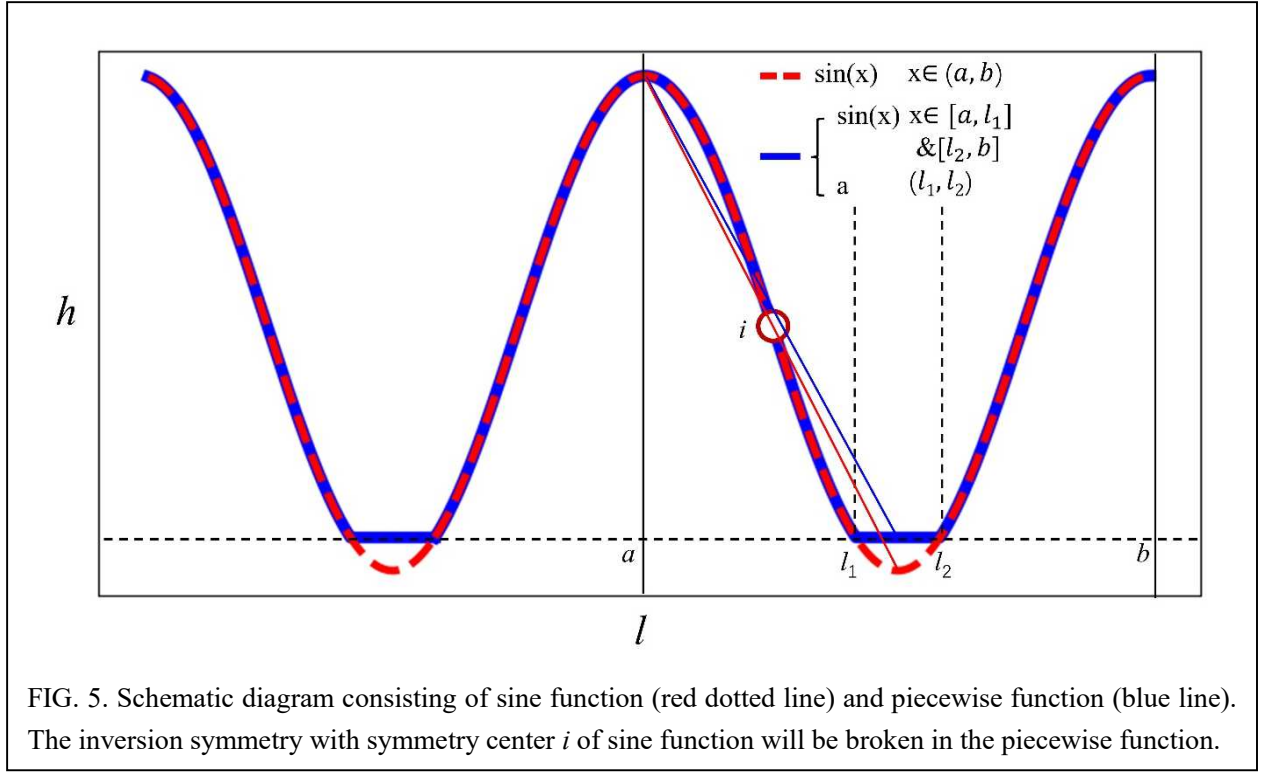

FIG. 5. Schematic diagram consisting of sine function (red dotted line) and piecewise function (blue line). The inversion symmetry with symmetry center *i* of sine function will be broken in the piecewise function.

In Fig. 2(i), we display the calculated results of DMI for 2D MnSe$_2$ with curvature. In 2D MnSe$_2$ structure, the global tendency of DMI is increasing as $c$ gradually increases, similar as that of 1D CrBr$_2$. In 2D MnSe$_2$, there are two DMI vectors when spin spiral vector $q$ propagates along directions of $\bar{\Gamma}-\bar{X}$ and $\bar{\Gamma}-\bar{Y}$ in the 2D BZ, in which the calculated values are in the range of 0-0.88 meV/f.u. and 0-26.29 meV/f.u., respectively. Obviously, compared to that $\bar{\Gamma}-\bar{Y}$ direction, the DMI strength along with $\bar{\Gamma}-\bar{X}$ direction is almost negligible. Thus, we focus on the DMI vector in the $\bar{\Gamma}-\bar{Y}$ direction below. The energy dispersion $E(q)$ and the DMI energy $\Delta E_{DMI}(q)$ associated with spin spiral vector $q$ for $c = 28.4\%$ and $c = 31.6\%$ are shown in Figs. 2(e)-2(h). The extracted DMI parameters $D$ by taking the slope of dispersion in the vicinity of ground state are 23.4 and 26.3 meV/f.u., respectively. For convenience in comparison, we also convert the total DMI per supercell to per magnetic atom as the effective DMI $D_{eff}$ (Fig. 1(g) and Fig. 2(i)). The $D_{eff}$, 0.21 and 0.24 meV/atom in 2D MnSe$_2$ induced by curvature with $c = 28.4\%$ and $c = 31.6\%$, which are comparable to those of Co/Ru(0001) [51] or Co/graphene interfaces [16,17], are highly possible to stabilize magnetic chiral spin textures in 2D magnets with long-range magnetic orderings.

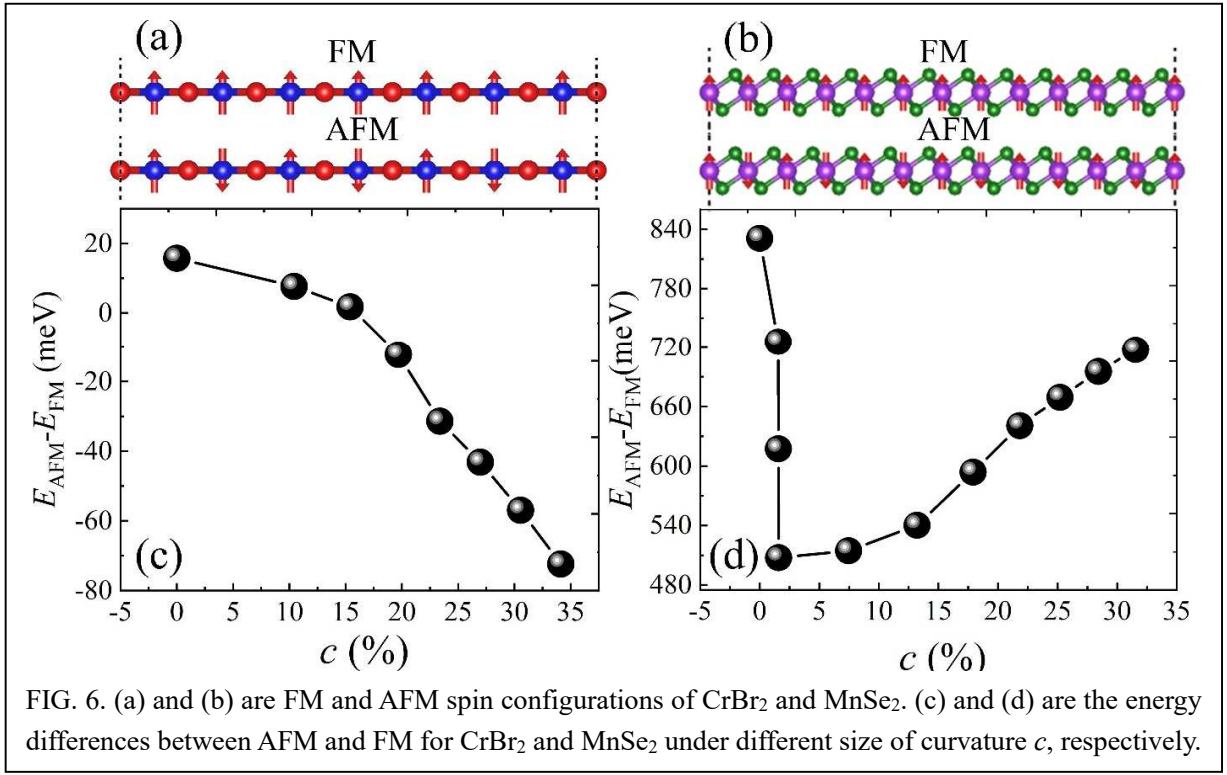

FIG. 6. (a) and (b) are FM and AFM spin configurations of $CrBr_2$ and $MnSe_2$. (c) and (d) are the energy differences between AFM and FM for $CrBr_2$ and $MnSe_2$ under different size of curvature $c$, respectively.

*Discussion.*—To further understand the variation of curvature-induced DMI above, the DMI effect of curved structure can be modelled as the sum of each M-X-M trimer, where M and X represent the magnetic and nonmagnetic atoms, respectively. Based on three-site Fert-Lévy model [52], the hoping of spins between two magnetic atoms (*i* and *j*) can induce DMI through a nonmagnetic atom (*l*) with SOC. The theoretically formulism of antisymmetric DMI can be expressed as:

$$\vec{D}_{ijl}(\vec{R}_{li}, \vec{R}_{lj}, \vec{R}_{ij}) = -V_{\text{soc}} \frac{(\vec{R}_{li} \cdot \vec{R}_{lj})(\vec{R}_{li} \times \vec{R}_{lj})}{|\vec{R}_{li}|^3 |\vec{R}_{lj}|^3 |\vec{R}_{ij}|}, \quad (1)$$

where $\vec{R}_{li}$, $\vec{R}_{lj}$, and $\vec{R}_{ij}$ are the corresponding distance vectors between atoms in the three-site model. $V_{\text{soc}}$ is a SOC-governed material parameter, its expressions can be written as $[135\pi\lambda_d \Gamma^2 (\sin(Z_d\pi/10)))/(32k_F^3 E_F^2)]$, in which $\lambda_d$, $k_F$, $E_F$, $Z_d$ are the SOC constant, Fermi vector, Fermi energy and the number of *d* electrons, respectively. $\Gamma$ is the interaction parameter between the localized spins and the spins of conduction electrons. In a curved magnet, the total DMI can be obtained by summing over all M-X-M trimers and we can further simplify the above Eq. (1) as:

$$\vec{D} = -V_{soc} \sum_{i=1}^{n} \frac{2*\sin(2\theta_i)\sin^2(\frac{1}{2}\theta_i)}{a_i^3}, \qquad (2)$$

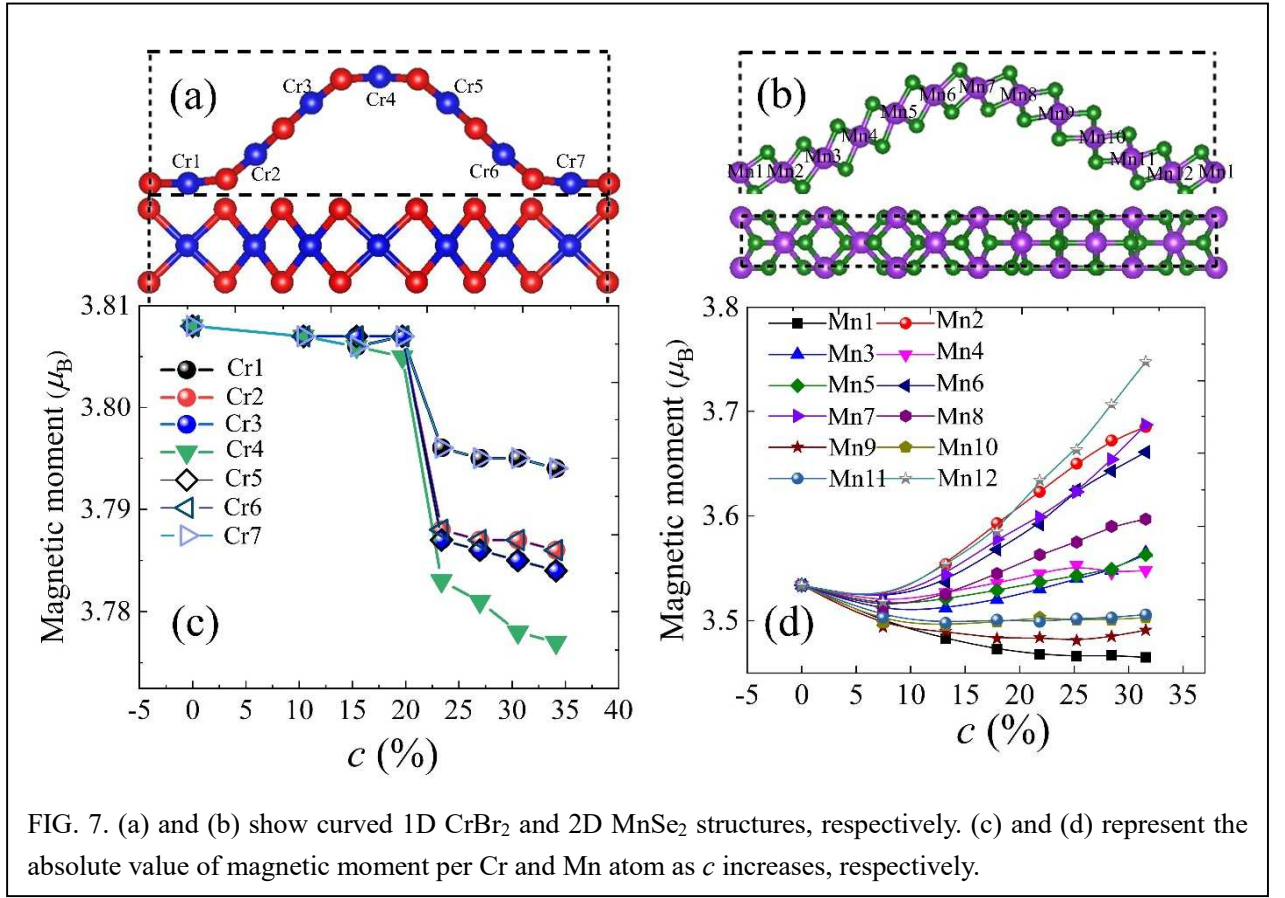

FIG. 7. (a) and (b) show curved 1D $CrBr_2$ and 2D $MnSe_2$ structures, respectively. (c) and (d) represent the absolute value of magnetic moment per Cr and Mn atom as $c$ increases, respectively.

where $i$ is the index of summation, $\theta_i$ and $a_i$ represent bond angle of each M-X-M trimer and bond length between the nearest neighbor M atoms in whole curved structure, respectively. In order to describe curvature dependence of DMI in Fert-Lévy model, we assume that $V_{soc}$ is a constant for a given system. Then we can provide a quantitative understanding for curvature-induced DMI based on Eq. (2). Insets $i$ in Figs. 3(a) and 3(b) present the schematic structures with inversion symmetry, which will lead to the offset of total DMI. However, the inversion symmetry breaking arising from curvature, will induce an uncompensated DMI in the noncentrosymmetric three-site interaction (insets $j$ in Figs. 3(a) and 3(b)). For 1D $CrBr_2$, in order to build up the relationship between curvature and DMI, the calculated $\theta_i$ and $a_i$ of each Cr-Br-Cr trimer are brought into the Eq. (2). The result shows that $D/2V_{soc}$ increases with increasing size of curvature $c$, as is shown in Fig. 3(a). Similar to that of 1D $CrBr_2$, we can determine the $\theta_i$ and $a_i$ of each Mn-Se-Mn trimer in curved 2D $MnSe_2$ and one can see that $D/2V_{soc}$ increases as size of curvature $c$ increases when $\theta_i$ and $a_i$ are brought into Eq. (2) as shown in Fig. 3(b). These results are in line with given results from first-principles calculations. More interestingly, notice that the whole curve structure is composed of several similar M-X-M trimers in our studied system, and we find that $\theta_i$ and $a_i$ of each M-X-M trimer present a small tendency of change as a whole in curved structure with increasing size of curvature $c$. Therefore, we propose a simple model, in which average bond angle $\bar{\theta}$ of Cr-Br-Cr (Mn-Se-Mn) triples and average bond length $\bar{a}$ of the nearest neighbor Cr (Mn) atoms in whole curve magnet are described as key parameters to further examine the relationship between curvature and DMI. We plot the dependences of $\bar{\theta}$ and $\bar{a}$ with respect to size of curvature $c$ as shown in Figs. 8(a) and 8(b). We can further simplify Eq. (2) as:

$$\vec{D} = -V_{\text{soc}} \frac{2*\sin(2\bar{\theta})\sin^2(\frac{1}{2}\bar{\theta})}{\bar{a}^3}. \qquad (3)$$

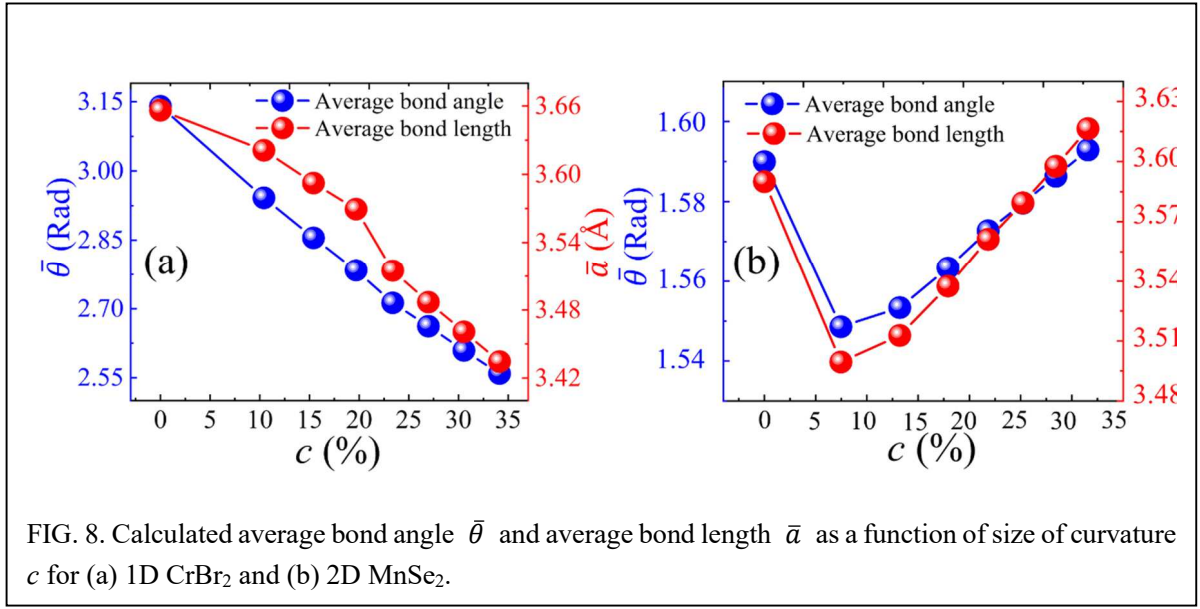

FIG. 8. Calculated average bond angle $\bar{\theta}$ and average bond length $\bar{a}$ as a function of size of curvature $c$ for (a) 1D CrBr$_2$ and (b) 2D MnSe$_2$.

Overall, one can see that $D/2V_{\text{soc}}$ increases for both 1D CrBr$_2$ and 2D MnSe$_2$ with increasing size of curvature $c$ when $\bar{\theta}$ and $\bar{a}$ are brought into Eq. (3), as shown in Figs. 9(a) and 9(b). This is also consistent with given results from first-principles calculations. Thus, our findings based on both first-principles calculations and Fert-Lévy model demonstrate that curvature is an effective way to realized DMI in 1D CrBr$_2$ and 2D MnSe$_2$.

*Realization of skyrmions in 2D MnSe$_2$.*—To explore the formation of such chiral spin textures in curved 2D MnSe$_2$, we apply atomistic spin model simulations [53] for the magnetization dynamics simulation based on magnetic parameters with size of curvature $c$ of 31.6% from first-principles calculations. To obtain final spin configurations, the atomistic

Landau-Lifshitz-Gilbert (LLG) equation is given:

$$\frac{\partial S_i}{\partial t} = -\frac{\gamma}{(1+\lambda^2)}\left[\mathbf{S}_i \times H_{eff}^i + \lambda \mathbf{S}_i \times \left(\mathbf{S}_i \times H_{eff}^i\right)\right], \qquad (4)$$

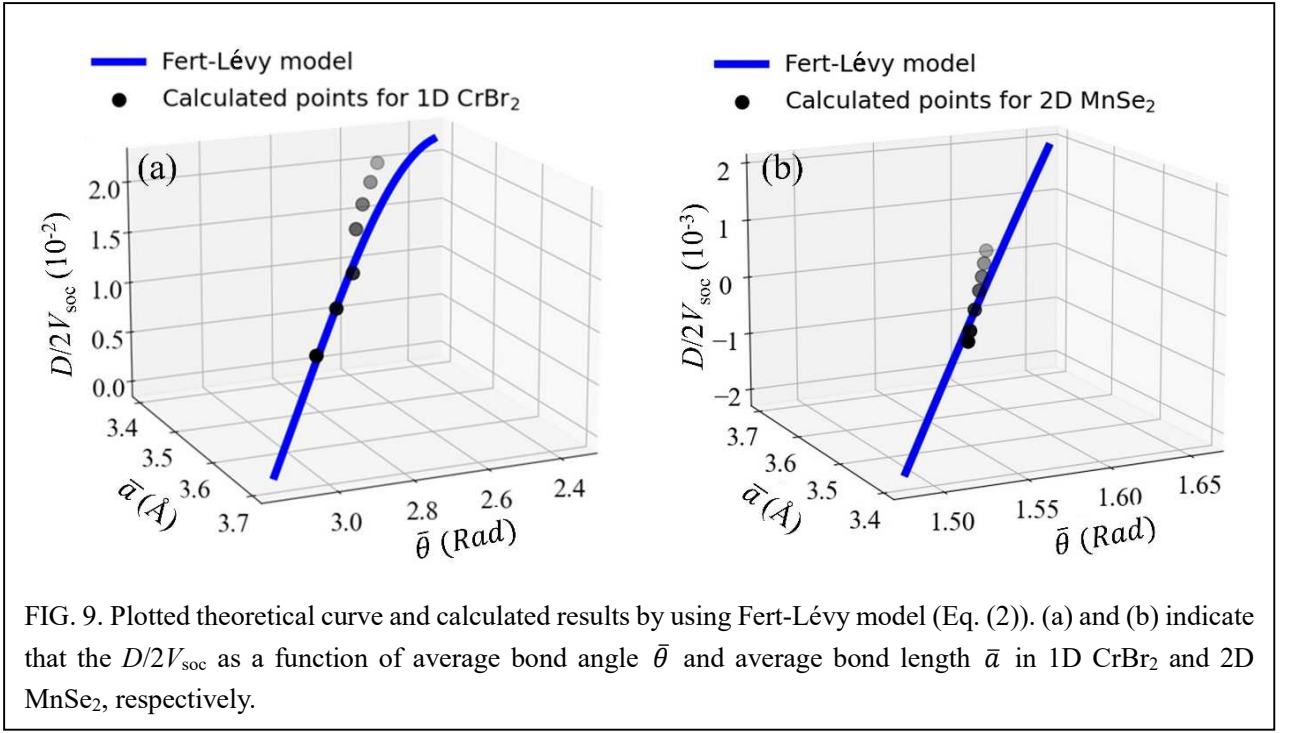

FIG. 9. Plotted theoretical curve and calculated results by using Fert-Lévy model (Eq. (2)). (a) and (b) indicate that the $D/2V_{soc}$ as a function of average bond angle $\bar{\theta}$ and average bond length $\bar{a}$ in 1D CrBr$_2$ and 2D MnSe$_2$, respectively.

Here $\gamma$ is the gyromagnetic ratio, $\lambda$ is the Gilbert damping constant and $\mathbf{S}_i$ is the unit vector, which is defined as $\frac{\mu_i}{|\mu_i|}$, where $\mu_i$ represents the magnetic moment of ith magnetic atom. $H_{eff}^i$ is the effective magnetic field on each $\mathbf{S}_i$ and can be obtained by following equation: $H_{eff}^i = -\frac{1}{\mu_i}\frac{\partial H}{\partial S_i}$, in which $H$ represents the total spin Hamiltonian. In simulations, we set a 60-nm width square with periodic boundary condition from an initial random state to obtain final spin textures. These parameters including effective exchange constant $J_{eff}$, effective DMI parameter $D_{eff}$, magnetic anisotropy $K$, dipole-dipole interaction $K_{Dip}$ and magnetic moment of Mn atoms are 14.94 meV, 0.24 meV, 0.17 meV, -0.176 meV and 3.58 $\mu_B$, respectively, and the methods to calculate $J_{eff}$, $K$, $D_{eff}$ and $K_{Dip}$ are described in the supplementary material. In addition, we apply the theoretical formula $\bar{R} = \pi D\sqrt{\frac{A}{16AK^2-\pi^2D^2K}}$ [54] to predict the radius $\bar{R}$ of skyrmion, which is close to 20 nm. We can find that the size of chiral structure is much larger than the ripple wavelength. Thus, we used the flat crystal structure to simulate our chiral magnetic structures in the atomic magnetic simulations. In simulations, Figure 4 shows the spin textures of curved MnSe$_2$ in $c$ = 31.6%. Interestingly, magnetic skyrmion spin texture with diameter of around 20 nm emerges under zero magnetic field as shown in Figs. 4(a) and 4(b). The calculated topological charge Q is -1 for skyrmion by using the formula $Q = \frac{1}{4\pi}\int \mathbf{S} \cdot (\partial_x \mathbf{S} \times \partial_y \mathbf{S})dxdy$, where $\mathbf{S}$ represents the unit vector, x and y are the coordinates.

*Conclusion.*—In summary, we obtain sizeable DMI in both 1D CrBr$_2$ and 2D MnSe$_2$ by using ripple structures with different curvatures via using first-principles calculations. We unveil that quite large DMI can be induced in curved structure, in which breaking of spatial inversion plays an important role. In addition, with the help of three-site Fert-Lévy model, the DMI dependence on the magnitude of curvature is successfully interpreted. Via atomistic spin model simulations, we further uncover that magnetic skyrmions can be induced without external field in curved 2D MnSe$_2$. Considering the experimental progress in 2D magnets, such as CrGeTe$_3$, CrI$_3$ and Fe$_3$GeTe$_2$, etc., our results open up new opportunities for development of spintronics.

*Note added.* After submission of this work, one article appeared on complimentary studies discussing curvature-induced magnetic parameters in curved magnets CrI$_3$ [55].

*Acknowledgments.* This work was supported by National Natural Science Foundation of China (Grants No. 11874059 and No. 12174405), Key Research Program of Frontier Sciences, CAS (Grant No. ZDBS-LY-7021), Ningbo Key Scientific and Technological Project (Grant No. 2021000215), Pioneer and Leading Goose R&D Program of Zhejiang Province under Grant No. 2022C01053, Zhejiang Provincial Natural Science Foundation (Grant No. LR19A040002), and Beijing National Laboratory for Condensed Matter Physics (Grant No. 2021000123).

**APPENDIX: Calculations of magnetic parameters**

**1. Dzyaloshinskii-Moriya interaction (DMI) *D***

DMI is beneficial to form the non-collinear magnetic textures, in which the neighboring spins tends to align perpendicular to each other. The energy term can be described as:

$$E_{DM} = \sum_{i,j} \mathbf{D}_{ij} \cdot (\mathbf{S}_i \times \mathbf{S}_j), \qquad (2)$$

where $D_{ij}$ represents the DMI vector, $S_i$ and $S_j$ are the spins of nearest neighbor magnetic atoms. We adopt the qSO method based on generalized Bloch theorem and treat SOC within the first-order perturbation theory [26,56,57], in which the total energy $E(q)$ as the function of spin-spiral vector $q$ is calculated along the high symmetry direction of $\bar{\Gamma}-\bar{X}$ and $\bar{\Gamma}-\bar{Y}$ in our studied system. To extract the DMI parameter, a Γ-centered $7 \times 1 \times 1$ for 1D $CrBr_2$ and $2 \times 18 \times 1$ k mesh for 2D $MnSe_2$ are considered to compute the spin spiral energy $E(q)$ in the interval of q from $-0.1\frac{2\pi}{a}$ to $0.1\frac{2\pi}{a}$ by a self-consistent way. The nonlinear spin spiral structure $m = (\sin(q \cdot r_i), 0, \cos(q \cdot r_i))$ rotates along the axis of spin spiral parallel the y axis, where $q$ represents the spin spiral vector and $r_i$ is the location of $i$th atom (When the rotation axis is x axis, the nonlinear spin spiral structure $m = (0, \sin(q \cdot r_i), \cos(q \cdot r_i))$). Then, DMI energy can be written as:

$$E_{DMI}(q) = \sum_{i,j} D_y \sin(q \cdot r_{ij}), \quad (3)$$

where $D_y$ represents the DM vector perpendicular to the propagation direction of spin spirals, and $r_{ij}$ is the unit vector between sites $i$ and $j$. Finally, the DMI parameter can be obtained by taking the slope of $\Delta E_{DMI}(q)$ with respect to spin spiral vector $q$, where $\Delta E_{DMI}(q)$ is the energy difference between $E(q)$ and $E(-q)$. Furthermore, final form of DMI energy can be simplified as:

$$\Delta E_{DMI}(q) = E(q) - E(-q) = 2\sum_{i,j} D_y \sin(q \cdot r_{ij}), \quad (4)$$

For small $q = (q,0,0)$ vector, we have the relationships:

$$\Delta E_{DM}(q) = 2q \sum_{i,j} D_{y0} r_{ij,x} = Dq, \quad (5)$$

where

$$D = 2\sum_{i,j} D_{y0} r_{ij,x}. \quad (6)$$

Here, the direction of DM vector depends on the propagation direction of spin spiral $q$ of the whole Brillouin zone.

## 2. Effective DMI value $D_{eff}$

In our calculations, we convert the total DMI $D$ per supercell to per magnetic atom as the effective $D_{eff}$. For 1D $CrBr_2$, $D = 14\pi D_{eff}$; For 2D $MnSe_2$, $D = 36\pi D_{eff}$.

### 3. Magnetocrystalline anisotropy energy *K*

*K* is defined as the energy difference between in-plane (100) and out-of-plane (001) magnetized axis:

$$K = E_{100} - E_{001} \quad (7).$$

### 4. Effective exchange coupling constant *J*$_{eff}$

As is shown in Fig. 6(b), we set different magnetic configurations FM and AFM for MnSe$_2$. We consider the Heisenberg model on a hexagonal lattice,

$$H = \sum_{ij} J_{eff} \vec{S}_i \cdot \vec{S}_j \quad (8)$$

Then effective exchange coupling constant can be solved by following formula:

$$E_{FM} = -\frac{1}{2} * 12 * (6J_{eff}) + E_{other}, \quad (9)$$

$$E_{AFM} = -\frac{1}{2} * 12 * (-2J_{eff}) + E_{other}, \quad (10)$$

$$J_{eff} = \frac{E_{AFM} - E_{FM}}{48}, \quad (11)$$

where the positive or negative value corresponds to FM or AFM coupling.

### 5. Dipole-dipole contribution to *K*$_{Dip}$

The dipole-dipole interaction contribution to *K* is the sum of all magnetostatic dipole-dipole interactions up to infinity [58,59] and is calculated by following formula:

$$E_{dip} = -\frac{1}{2}\left\{\frac{\mu_0}{4\pi V_{u.c.}}\right\} \sum_{i,j=1}^{N} \frac{(\boldsymbol{m}_i \cdot \boldsymbol{m}_j)r_{ij}^2 - 3*(\boldsymbol{m}_i \cdot \boldsymbol{r}_{ij})(\boldsymbol{m}_j \cdot \boldsymbol{r}_{ij})}{r_{ij}^5}, \quad (12)$$

where $\boldsymbol{m}_i$ and $\boldsymbol{m}_j$ represent the unit vector of magnetization at position $\boldsymbol{r}_i$ and $\boldsymbol{r}_j$, $\boldsymbol{r}_{ij}$ is the unit vector between site *i* and *j*. In calculations, we choose a $150 \times 150 \times 1$ supercell to obtain the dipole-dipole interaction between two magnetic atoms in 2D MnSe$_2$.